\documentclass[10pt, conference]{IEEEtran}
\IEEEoverridecommandlockouts
\usepackage{cite}
\usepackage{amsmath,amssymb,amsfonts}
\usepackage{algorithmic}
\usepackage{graphicx}
\usepackage{textcomp}
\usepackage[dvipsnames]{xcolor}
\usepackage{url}

\usepackage{booktabs}
\usepackage{hyperref}
\usepackage{adjustbox}
\usepackage{tcolorbox}
\usepackage{comment}
\usepackage{multirow}
\usepackage{xtab}
\usepackage{soul}
\usepackage{verbatim}
\usepackage{fontawesome}
\definecolor{darkpastelgreen}{rgb}{0.01, 0.75, 0.24}
\definecolor{ao}{rgb}{0.0, 0.5, 0.0}
\definecolor{applegreen}{rgb}{0.55, 0.71, 0.0}
\definecolor{forestgreen}{rgb}{0.13, 0.55, 0.13}
\definecolor{green(ryb)}{rgb}{0.4, 0.69, 0.2}
 \definecolor{aqua}{rgb}{0.0, 1.0, 1.0}

\newcommand{\rqone}{What Challenges are Associated with Reusing PTMs in the HF Community?}
\newcommand{\rqtwo}{What are the Benefits of the HF Community for Reusing PTMs?}
\newcommand{\rqthree}{What are the Trends in PTM Reuse within the HF Community?}
\newcommand{\hf}{HF }

\def\BibTeX{{\rm B\kern-.05em{\sc i\kern-.025em b}\kern-.08em
    T\kern-.1667em\lower.7ex\hbox{E}\kern-.125emX}}

\begin{document}

\title{Deep Learning Model Reuse in the HuggingFace Community:
Challenges, Benefit and Trends\\
\thanks{This work was supported by: Fonds de Recherche du
Québec (FRQ), the Canadian Institute for Advanced Research
(CIFAR) as well as the DEEL project CRDPJ 537462-18
funded by the Natural Sciences and Engineering Research
Council of Canada (NSERC) and the Consortium for Research
and Innovation in Aerospace in Québec (CRIAQ), together
with its industrial partners Thales Canada inc, Bell Textron
Canada Limited, CAE inc and Bombardier inc.}
}

\author{\IEEEauthorblockN{Mina Taraghi, Gianolli Dorcelus, Armstrong Foundjem, Florian Tambon, Foutse Khomh}
\IEEEauthorblockA{\textit{SWAT Lab., Polytechnique Montréal}\\
Montréal, QC, Canada \\
\{mina.taraghi, gianolli.dorcelus, a.foundjem, florian-2.tambon, foutse.khomh\}@polymtl.ca}}

\maketitle
\thispagestyle{plain}
\pagestyle{plain}

\begin{abstract}
The ubiquity of large-scale Pre-Trained Models (PTMs) is on the rise, sparking interest in model hubs, and dedicated platforms for hosting PTMs. Despite this trend, a comprehensive exploration of the challenges that users encounter and how the community leverages PTMs remains lacking. To address this gap, we conducted an extensive mixed-methods empirical study by focusing on discussion forums and the model hub of HuggingFace, the largest public model hub.
Based on our qualitative analysis, we present a taxonomy of the challenges and benefits associated with PTM reuse within this community. We then conduct a quantitative study to track model-type trends and model documentation evolution over time. Our findings highlight prevalent challenges such as limited guidance for beginner users, struggles with model output comprehensibility in training or inference, and a lack of model understanding. 
We also identified interesting trends among models where some models maintain high upload rates despite a decline in topics related to them.
Additionally, we found that despite the introduction of model documentation tools, its quantity has not increased over time, leading to difficulties in model comprehension and selection among users.
Our study sheds light on new challenges in reusing PTMs that were not reported before and we provide recommendations for various stakeholders involved in PTM reuse.
\end{abstract}


\section{Introduction}\label{sec:introduction}
Recently, Deep Neural Networks (DNNs) have been leveraged successfully in many domains to address real-world problems. Yet, training DNNs from scratch is usually very time-consuming, computationally expensive, and has a significant carbon footprint \cite{jiang2022empirical}\cite{lacoste2019quantifying}\cite{patterson2021carbon}. Consequently, practitioners are increasingly reusing and adapting Pre-Trained Models (PTMs) to their use cases instead of training new models from scratch. Pre-training Deep Learning (DL) models with huge amounts of data was first successfully applied to computer vision \cite{krizhevsky2012imagenet}\cite{oquab2014learning} and NLP tasks \cite{collobert2011natural}, but in recent years, PTMs have showcased exceptional performance across language understanding and generation tasks, often surpassing human-level results \cite{han2021pre}. These successes and progress in the NLP community have shifted the spotlight of AI research towards large-scale PTMs \cite{han2021pre}. PTMs are also extensively used by Software Engineering (SE) researchers \cite{hou2023large}\cite{luo2022prcbert}\cite{he2022ptm4tag} and developers \cite{murali2023codecompose}.

Following this surge in model reuse, and to facilitate its usage, ``Model Hubs'' have been created. Model hubs are platforms that host collections of PTMs and/or datasets that have been curated and categorized to be easily reused by users\cite{jiang2022empirical}. HuggingFace (HF) Hub is the biggest model hub on the internet \cite{jiang2023empirical} with more than 371K models as of October 2023\cite{huggingfaceModelHub}. 

\begin{figure*}[ht]
    \centering
    \includegraphics[width = 1\textwidth]{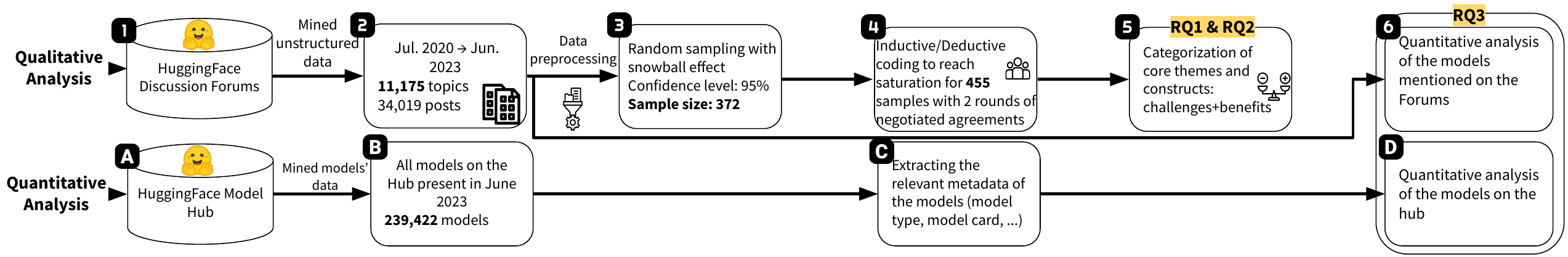}
    \caption{The mixed-methods methodology followed in the study.} 
    \label{fig:methodology}
    \vspace{-1.5em}
\end{figure*}

Recent studies have investigated challenges, bugs, and issues that arise when reusing PTMs from the model hubs
\cite{jiang2023empirical}\cite{pan2022empirical}\cite{davis2023reusing}, but their scope 
have been limited. Moreover, there is a lack of a comprehensive view of how the community leverages and affects PTMs reuse. Hence, in this study, we empirically examine how the HF community reuses PTM by analyzing discussions in the HF Forums \cite{huggingfaceHuggingFace}, the main platform used by the HF community for discussing a wide range of topics and concerns. Using a mixed-methods approach, we perform a large-scale study on the challenges, benefits, and trends in PTM reuse. First, we mine and perform qualitative analysis on the HF forums' discussion threads to understand the challenges that users face in reusing PTMs and the potential benefits that the community brings to PTM reuse. Next, we quantitatively investigate and compare the distribution of the types of models being discussed by the community and those uploaded and available on the hub to identify trends in PTMs reuse within the community. The overview of our methodology is shown in Figure \ref{fig:methodology}.

To guide our investigation, we define three Research Questions (RQ):

\begin{itemize}
    \item[\textbf{RQ1.}] What challenges are associated with reusing PTMs in the HF community?
    \item[\textbf{RQ2.}] What are the benefits of the HF community for reusing PTMs?
    \item[\textbf{RQ3.}] What are the trends in PTM reuse within the HF community?
\end{itemize}

As a result of our qualitative analysis, we recognize 17 categories (Table \ref{tab:challenges_and_benefits}) of challenges faced by the community that were further categorized into 47 subcategories. Among these, the challenges related to \textit{understanding models and their functionalities}, \textit{requesting a complete solution for a specific use case} and \textit{difficulty in interpreting the output of the model in training or inference} were the most prevalent challenges faced by the users which drastically hampers PTM reuse. On the other hand, we also found 6 categories of benefits associated with the community for PTM reuse. Our findings highlight the forums' pivotal role in fostering collaborations among diverse stakeholders. The solutions provided by the experts, and various types of collaboration among users and experts for effective problem-solving have also emerged as significant advantages within the HF community.
Our quantitative analysis results not only confirm the previous findings \cite{you2022ranking} suggesting that the model hubs are under-exploited, but also shows that the distribution of models provided on the HF model hub and the models discussed by the HF community is similar, and BERT, some BERT-based models (DistilBERT, RoBERTa), GPT2, T5, and BART are among the most popular ones, with BERT being the most discussed and most uploaded model type over time with a noticeable distance from others. We also find that despite the introduction of tools for documenting models, the quantity of model documentation has not improved over time. In light of our observations, we provide guidelines on PTM reuse that can be leveraged by the HF community, and to some extent, generalized to model hub communities.

Thus, we make the following contributions in this study:

\begin{itemize}
    \item[-]  We provide a taxonomy of challenges and benefits of PTM reuse from the point of view of the HF community.
    \item[-] We study and discuss existing trends in the HF community in PTM reuse.
    \item[-] We provide guidelines for the community towards the betterment of PTM reuse. While those guidelines mainly apply to the HF community, they should be helpful to some extent to other model hubs.
\end{itemize}

The remainder of the paper is structured as follows: In \textbf{Section \ref{sec:background}} we give a background on HF as a platform and the community. Next, we will explain the qualitative part of our study in \textbf{Section \ref{sec:qualitative}} and answer RQ1 and RQ2. The quantitative steps in the study are explained in \textbf{Section \ref{sec:quantitative}} where we answer RQ3. Finally, we will discuss the results and implications in \textbf{Section \ref{sec:discussion}}, before detailing in \textbf{Section \ref{sec:relatedwork}} the related works and in \textbf{Section \ref{sec:threats}} the threats to validity. Finally, we conclude the paper in \textbf{Section \ref{sec:conclusion}}.

\section{Background}\label{sec:background}
HuggingFace is a PTM reuse platform that provides open-source resources and services to the community. They provide several libraries for reusing PTMs, the most notable of which is \textit{HuggingFace Transformers}\cite{huggingfaceTransformers}. HF model hub (hereafter referred to as \textit{the hub}) is the biggest PTM hosting platform on the web and the only hub to be \emph{open}, among the most popular ones, which means there is no review of models uploaded on this hub; neither the model contributor nor the uploaded models have to be reviewed or verified by anyone else before being accessible to the public \cite{jiang2022empirical}. 
Apart from models, HF also provides a dataset hub \cite{huggingfaceDatasets} and a space hub \cite{huggingfaceSpaces} and different libraries to facilitate their usage with the models. HF also collaborates with proprietary companies such as  Amazon \cite{huggingfaceAmazonSagemaker}\cite{huggingfaceOptimumNeuron} and Gradio \cite{gradioGradioInterface} to provide resources for training, inference, deployment, and demoing of the models. Therefore the HF platform provides services and resources to support the entire PTM reuse life cycle. As of October 2023, HF hosts over 371K models, 73K datasets and 133K spaces (model demos).

HF used various platforms for collaboration, discussion, and troubleshooting. Prior to the launch of the forums in July 2020, the main discussions and troubleshooting took place in the issues section of the GitHub repositories of various HF libraries. Following the HF community's growth, a dedicated platform was needed for discussions and question-answering~\cite{githubUpdatingIssue}\cite{huggingfaceWhatKind}. Thus, HF established the Forums~\cite{huggingfaceHuggingFace} to facilitate collaboration within the open-source community and among users and to keep the GitHub issues devoted to ``bug reports'' and ``feature requests'' \cite{huggingfaceWhatKind}. 

HF forums are powered by \emph{Discourse}~\cite{Discourse} which is an open-source online tool that supports discussion topics organized in threads, different categories, tagging, etc. Each topic consists of one or more entries known as posts, which are either a request for help, a question, or a reply to a question. Also, the original author of a topic can mark a response as \emph{`accepted answer'} similarly to StackOverflow~\cite{stackoverflow}.

\section{Qualitative Analysis }\label{sec:qualitative}
In this section, we describe the motivation behind our qualitative analysis, the methodology of data collection and the steps taken to answer RQ1 and RQ2. The findings for these RQs are subsequently explained.

\subsection{Motivation} \label{subsec:qual-motivation}
The HF community is a collection of diverse stakeholders engaged in various socio-technical activities related to model reuse across different domains and tasks \cite{huggingfaceusers}\cite{huggingfaceOrganizations}. Given HuggingFace's mission to ``\textit{democratize good machine learning}'' \cite{huggingfaceAboutUs} by making PTMs accessible, understanding challenges and leveraging community benefits can aid HF (and other platform providers), model providers (e.g., OpenAI, Google, etc.,), and users. Insights from such a study can enhance PTM service providers' efficiency, improve user experience, and enable the creation of more efficient models for the public. This motivated us to conduct a study to uncover the challenges in PTM reuse in HF and to find out the benefits that communities such as HF bring to the PTM reuse.

\subsection{Methodology}\label{subsec:methodology} 
\subsubsection{Data Collection}\label{subsubsec:data_collection}

The HF Forums started with 8 categories in 2020 and by the time of data collection in June 2023, there were 26 categories present. These categories are not of the same nature (e.g., levels of expertise, different external platform services, HF libraries, etc.) and while each discussion can be started in only one category, it can be related to multiple categories (e.g., a ``Beginner'' question asked about a ``Transformer'' ``model'').

\textbf{Data Mining:} As mentioned in Section \ref{sec:background}, \hf forums \cite{huggingfaceHuggingFace} has been the principal platform for discussion and troubleshooting of the users. Therefore, the HF forums were used as the main subject for our qualitative analysis. For collecting the data to construct the dataset, we mined all the publicly available forum topics using Grimoirelab's Perceval library \cite{duenas2018perceval}. The mined data spanned over a period of 35 months, starting from the forum's inception with the first topic created on 7 July 2020 up to the date of mining on 26 June 2023.

Using the Perceval library, we parsed the Forums' URL \cite{huggingfaceHuggingFace} with the desired timeframe as a command option to extract the topics' data in a JSON format. Forums' topics are arranged per page, i.e., a page contains one or more posts by one or more users (i.e., authors of posts). We further parsed the JSON file to extract the metadata (containing userID, postID, etc.), topics, and posts. Additionally, we linked each post to the user, topic, and category.

In total,  \textbf{11,175} topics containing \textbf{34,019} posts were mined and parsed. We then crawled the webpage for each topic and converted each page to a PDF file, forming a raw, unstructured textual dataset (Figure \ref{fig:methodology}, Step 2), which we then analyzed qualitatively in the next step. In particular, a HF Forums' structure is organized as the following URL format: \\\textit{\url{https://discuss.huggingface.co/t/TITLE/TOPIC/POST}}.
\\Finally, using a 95\% confidence level with a 5\% margin of error, we randomly selected \textbf{372} topics from the population of 11,175 to be processed in the next step (Figure \ref{fig:methodology}, Step 3).

\subsubsection{Qualitative Analysis}\label{subsubsec:extracting_themes}
We followed a methodology similar to Foundjem et al. in~\cite{mixed-method22} by using thematic analysis to identify key benefits and challenges that arise in the HF community when reusing models. We used content analysis to understand the prevalence of these themes discussed by practitioners and users in the HF community. We then constructed a hierarchy of themes and tracked their popularity based on the number of occurrence~\cite{nardone2023video}. Since the dataset consists of unstructured and textual discussions with varying numbers of posts and no predefined taxonomy or categorization for this data, we used an inductive coding approach to code contents in the text at the sentence and paragraph levels \cite{ferreira21,mohanani22,ghorbani2023autonomy}. 

\textbf{Inductive/Deductive Coding:} For this step we followed a similar methodology to Foundjem et al. in \cite{foundjem2021onboarding}. 
Initially, in the inductive coding process~\cite{ferreira21}, the first three authors (raters) were each assigned 15\% (56 topics) of the 372 samples to code independently, by using descriptive labels of their own. The raters then met after coding the first eight topics (15\% of the 56 topics) to discuss their codes and resolve disagreements, i.e., a negotiated agreement~\cite{mixed-method22}, and this process initially built our codebook. Another round of negotiations was done at the end of inductive coding (after coding 56 samples). Next, two other raters continued the coding process by deductively~\cite{ferreira21} applying the agreed codes to new instances of the topics. As the deductive coding process continued, new labels that emerged were added to the codebook until no new theme emerged. At this point, we reached saturation~\cite{ferreira21,mixed-method22} and stopped the coding process, leading us to \textbf{455} topics. Then, we calculated the Inter-Rater Reliability (IRR) using Cohen Kappa ($\kappa$)~\cite{mixed-method22}, which yielded $\kappa$ = 0.74 in the first round. We resolved disagreements and in the second round had $\kappa$ = 0.88. Lastly, we merged codes in agreed categories, resulting in a perfect coding agreement of $\kappa$ = 1 (a perfect score) in the last round.

As part of our process, we explored links within each discussion topic that led to other topics or external sources such as GitHub issues, documentation, or StackOverflow posts. Studying the content of these links helped us gain a better understanding of the discussions and arguments driving the challenges and benefits within the HF community, allowing us to better contextualize our labels and insights. In particular, we obtained \textbf{1237} low-level codes. For further analysis, we exported these codes to Miro \cite{miro} which is a collaborative visual workspace for managing ideas \cite{AboutMiro}.

\textbf{Thematic/Content Analysis and Data Visualization:} 
To facilitate the thematic analysis, the first author initially grouped similar extracted codes into common categories, forming a cluster of different (sub-)categories and hierarchically constructed the first affinity diagram using the Miro board~\cite{miro}; details of our codes and categories are available in our replication package~\cite{replication}. Once the first version of the diagram was developed, the first three authors met to discuss refinement. They then met repeatedly, deliberated, and refined the categories. Based on those discussions, the first author refined the diagram (Figure \ref{fig:methodology}, Step 5). 

After conducting a thematic analysis, we identified \textbf{17} themes related to challenges and \textbf{6} themes related to benefits that emerged during the labelling process of sub-categories and categorization, in total, \textbf{23} themes. To gauge the prevalence of each theme, we used content analysis to measure the strength of their relation to the challenges and benefits, i.e., we counted the frequency of occurrence of a given theme/concept against the entire body of emerged themes. Then we sorted the occurrences in descending order with the absolute value representing the importance/strength of that theme in relationship to the discussion. Table~\ref{tab:challenges_and_benefits} shows the outcome of both the thematic and content analysis.

\begin{table}[!ht] 
    \centering
     \caption{Challenges and Benefits associated with PTM reuse}
     \vspace{-1em}
     \label{tab:challenges_and_benefits}
    \begin{tabular}{lcr}
    \toprule
        \textbf{Challenges} & \textbf{Count} & \textbf{(\%)} \\ \midrule
        \textbf{(C1)}  Model usage and understanding & 106 & 14.38 \\ 
        \textbf{(C2)}  Training pipeline & 88 & 11.94 \\ 
        \textbf{(C3)}  Memory \& performance & 80 & 10.85 \\ 
        \textbf{(C4)}  External platforms \& libraries & 67 & 9.09 \\ 
        \textbf{(C5)}  Tutorial \& documentation & 60 & 8.14 \\ 
        \textbf{(C6)}  Specific solution need & 56 & 7.60 \\ 
        \textbf{(C7)}  Dataset acquisition \& usage & 51 & 6.92 \\ 
        \textbf{(C8)}  Incomprehensible output & 51 & 6.92 \\ 
        \textbf{(C9)}  HF libraries' features & 46 & 6.24 \\ 
        \textbf{(C10)}  API usage & 31 & 4.21 \\ 
        \textbf{(C11)}  Inference \& deployment & 28 & 3.80 \\ 
        \textbf{(C12)}  Website \& pricing & 26 & 3.53 \\ 
        \textbf{(C13)}  Spaces & 16 & 2.17 \\ 
        \textbf{(C14)}  Onboarding & 9 & 1.22 \\ 
        \textbf{(C15)}  Privacy \& security & 9 & 1.22 \\ 
        \textbf{(C16)}  Other issues & 7 & 0.95 \\ 
        \textbf{(C17)}  Lack of support by HF & 6 & 0.81 \\  
        \midrule
        \midrule
        \textbf{Benefits} & \textbf{Count} & \textbf{(\%)}  \\ \midrule
        \textbf{(B1)} Expert opinion and Solutions & 262 & 50.98  \\ 
        \textbf{(B2)} Collaboration to Solve Issues & 141 & 27.43 \\ 
        \textbf{(B3)} Acknowledgement of help & 76 & 14.78 \\ 
        \textbf{(B4)} Call for collaboration & 15 & 2.92 \\ 
        \textbf{(B5)} Announcements & 12 & 2.33 \\
        \textbf{(B6)} Users from other domains & 8 & 1.56 \\ 
        \bottomrule
    \end{tabular}
    \vspace{-2em}
\end{table}

\textbf{Root Cause Analysis on Reported Issues:} To further investigate the provenance of the identified issues, we followed the Root Cause Analysis (RCA) method similar to Zhang et al. \cite{tianyi2019RCA}. We extracted the topics that had accepted answers to analyze the root causes of the issues. Since the ratio of accepted answers is low for the \hf forums (0.09 for the whole dataset and 0.11 for our selected samples), we also went through all the discussions in the sample and marked all the discussions in which there was a verbal indication of the answer being accepted by the user who asked the question.

Next, the first two authors went through all the discussions with either form of an accepted answer to recognize the root cause of each issue. Also, the authors grouped the recognized root causes into a hierarchical structure using an affinity diagram, similar to the last step (also accessible in our Miro board \cite{miro}).
\subsection{RQ1. \rqone}\label{subsec:RQ1}
\subsubsection{Taxonomy of Challenges} \label{subsubsec:RQ1-taxonomy}
We grouped the issues faced by the HF users into 17 categories and 47 subcategories. The overview of the categories can be found in Table \ref{tab:challenges_and_benefits}. We will explain the most notable categories and their corresponding subcategories in the rest.

The most prominent category of challenges that HF users face, directly concerns \textit{using and understanding models}, \textbf{C1} (14.38\%). The biggest subcategory of this group of issues manifests when users are having difficulty \textit{understanding the model definition or functionality} (6.65\%). 
The rest of this category belongs to issues people face in making their \textit{custom models} work, \textit{converting, evaluating and comparing, loading and saving, and configuring models}. 

The next biggest category of issues belongs to the \textit{training pipeline} and various issues that they face when they want to \textit{pre-train, train, or fine-tune a model}, \textbf{C2} (11.94\%). \textit{Tokenization} issues are also included in this group. 

The \textit{memory and performance}, \textbf{C3} also make up a big part of the challenges (10.85\%). Many of these issues are related to handling the \textit{large files} of the model weights and using \textit{multiple GPUs for multiprocessing} during training and inference. 

Next, there is an interesting category involving the challenges users face when using \textit{platforms and libraries other than HF} (\textbf{C4}) with it, accounting for 9.09\% of the cases. Two partnerships of HF, \textit{Amazon (all services)} (1.76\%) and \textit{Gradio} (1.09\%) are at the top of the list of the platforms with which users face issues when reusing models.

Next, we encounter challenges associated with using or seeking \textit{documentation and tutorials}, \textbf{C5} comprising 8.14\% of the total. Within this category, 4.61\% of the issues are tied to the \textit{official tutorials, examples, and notebooks} provided by HF. Additionally, 3.53\% of cases in the dataset involve users highlighting \textit{deficiencies in the documentation}.

The largest category without any subcategories is \textit{specific solution need}, \textbf{C6} (7.6\%). This category focuses on questions where the user presents a particular problem or task with specific conditions, seeking assistance on how to approach it. 
Next, we observe issues related to \textit{dataset acquisition and usage}, \textbf{C7} constituting 6.92\% of the total challenges. This category encompasses specific subcategories such as \textit{creating and customizing datasets} (1.9\%) and \textit{preparing data} for training or inference (1.49\%).

The next large category with no subcategories, \textit{incomprehensible outputs}, \textbf{C8} relates to the confusion the users face when they cannot comprehend or justify the output produced by the model in training or inference (6.92\%). 

The last large category of issues belongs to the requests and questions that are related to the \textit{HF libraries' features}, \textbf{C9} (6.24\%). This category is comprised of the questions that users ask about the \textit{implementation of HF source code} and about the availability of features (4.88\%) and \textit{requests for a specific feature or model} (1.36\%).

Other categories of challenges, each of which is responsible for less than 5\% of the total challenges detected, are \textit{API usage (\textbf{C10}), inference and deployment (\textbf{C11}), website and product pricing (\textbf{C12}), spaces (\textbf{C13}), onboarding (\textbf{C14}), privacy and security (\textbf{C15}), and lack of support for services (\textbf{C17})}. The definition and examples for all these categories (and their subcategories) are available in our codebook included in the replication package \cite{replication}.
There were also a total of seven issues that we could not categorize in any of the categories above (\textit{Other issues}, \textbf{C16}). 

\subsubsection{Root Cause Analysis}\label{RQ1-RCA} For analyzing the root causes, we identified 21 issues for which there was a verbal affirmation by the user that the answer was accepted. Adding these 21 issues to the 51 answers that were officially marked as accepted answers, increased the ratio of accepted answers in the sample from 0.11 to 0.16. The results for the grouping of the root causes are listed in Table \ref{tab:rootcause}. \textit{User-related} and \textit{HF-related} causes make up 58.3\% and 33.3\% of the root causes, respectively. The categories of the HF-related issues comprise four types of lack of knowledge by the user, with the lack of knowledge about HF libraries, and HF APIs, responsible for roughly one-third or 31.9\% of all the root causes, followed by lack of programming and lack of theoretical knowledge about the subject of their question/issue. The HF platform
is responsible for three types of issues, \textit{lacking features}, \textit{lacking documentation for HF libraries and services}, and \textit{HF bugs} with 16.7\%, 12.5\%, and 4.2\% of all the root causes respectively. External platforms and libraries take the blame for 2.8\% of the analyzed issues. The issues here include an instance for Amazon where the mapping of files on Amazon Sagemaker caused confusion for several users and the expert points it out in \textit{T31171: ``This is indeed very confusing at first, but it stems from the fact that the S3 locations will be mapped to local folders in the file system of the training job instance''}\cite{amazonissue}. The other instance also is when the Gradio member admits to lacking feature by Gradio in \textit{T19446: ``We will support column width’s directly soon, but for now you can apply custom CSS and elem\_id’s''}\cite{gradioissue}.

\begin{table}[t]
\centering
\caption{\centering The Root Causes of Challenges of PTM Reuse}
\vspace{-1em}
\label{tab:rootcause}
\begin{tabular}{llcr}
\toprule
\textbf{Category}                                                             & \textbf{Subcategory}                                                                     & \textbf{Count} & \textbf{(\%)} \\ \hline
User-Related                                                                  & \begin{tabular}[c]{@{}l@{}}Lack of familiarity with \\ HF and its libraries\end{tabular} & 15             & 20.8               \\
                                                                              & \begin{tabular}[c]{@{}l@{}}Lack of practical and \\ programming knowledge\end{tabular}   & 12             & 16.7               \\
                                                                              & \begin{tabular}[c]{@{}l@{}}Lack of API \\ usage knowledge\end{tabular}                   & 8              & 11.1             \\
                                                                              & \begin{tabular}[c]{@{}l@{}}Lack of theoretical\\  knowledge\end{tabular}                 & 7              & 9.7               \\ \hline
HF-Related                                                                    & \begin{tabular}[c]{@{}l@{}}LackingFeature \\ by HF\end{tabular}                          & 12             & 16.7               \\
                                                                              & Lacking documentation                                                                    & 9              & 12.5              \\
                                                                              & HF bugs                                                                                  & 3              & 4.2               \\ \hline
\begin{tabular}[c]{@{}l@{}}External   Platforms \\ and Libraries\end{tabular} &                                                                                          & 2              & 2.8               \\ \hline
Other/Unknown                                                                 &                                                                                          & 4              & 5.5               \\ \bottomrule
\end{tabular}
\vspace{-2em}
\end{table}

\subsubsection{Comparison of Challenges based on the Expertise Level}\label{RQ1-expertise} 
\begin{figure*}[ht]
    \centering
    \includegraphics[width=\textwidth]{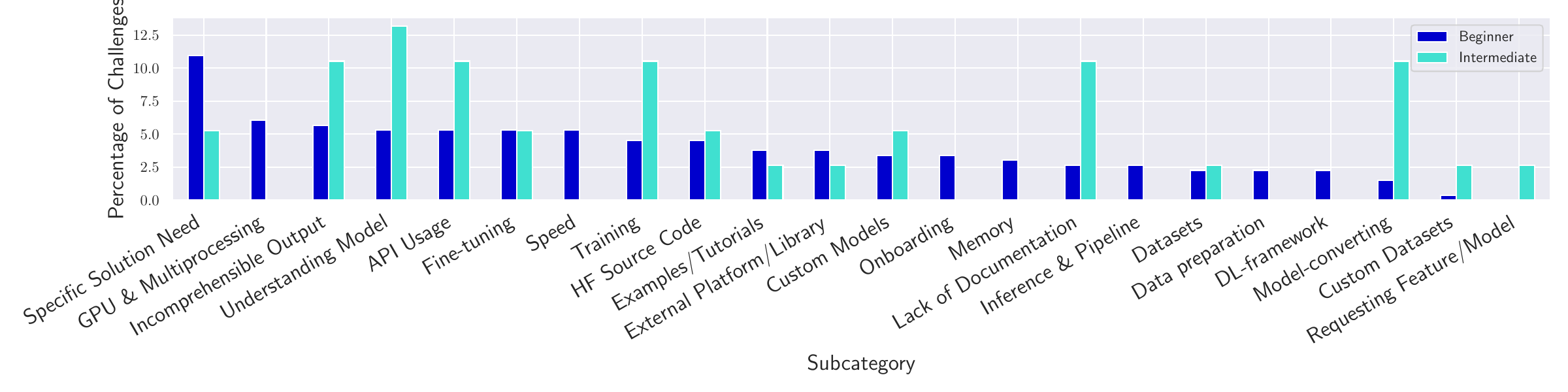}
    \caption{The comparison of the percentage of some subcategories of challenges with ``Beginner'' and ``Intermediate'' tags}
    \label{fig:beginner_inter}
    \vspace{-1em}
\end{figure*}

As mentioned in Section \ref{subsubsec:data_collection}, each discussion in the forum is posted in a certain category. Among these categories, there are two that are related to the expertise level of the user or the difficulty of their question, i.e., \emph{Beginners} and \emph{Intermediate}. By the time of mining, the largest category of discussions on the forums was ``Beginners'' (3.7K topics) and around 7.4 times the number of the ``Intermediate'' category topics (0.5K topics). To analyze the difference between the levels of expertise we also separated the topics for these two categories and compared the distribution of subcategories. The ratio of these two categories was also 7 to 1 in our dataset and in the codes extracted (Beginner to Intermediate). The ratio of subcategories (with a share of more than 2\%) of challenges for these two levels are reported in Figure \ref{fig:beginner_inter}.

One probable reason for this big difference between the share of \textit{beginner} and \textit{intermediate} challenges could be due to beginner users being less specific about the category of their question because of their lower knowledge, and prefer to post them in a rather general category, while intermediate users who are a bit more knowledgeable ask their questions in the category that is suitable for the question. For example the user in \textit{T35671} \cite{beginnerDataset} is asking a question about a translation task following the official tutorial for their custom dataset and expressing doubt on whether the issue is because of the dataset pre-processing: \textit{``It’s a problem with the dataset format right? What’s the right format, and how can I process it?''}. Nevertheless, they decided to ask it in the \textit{beginners} category.

As shown in Figure \ref{fig:beginner_inter}, the share of challenges related to custom models (5.3\%) and custom datasets (2.6\%) are larger for the intermediate users, and intermediate users are more concerned with requesting features/models (2.6\%) or asking about the implementation of features in HF libraries' source code (5.3\%). This shows that intermediate users are more involved with customizing their solutions and the features that are provided by the company compared to beginner users.
It is also worth noting that while a big part of the beginners come to the forums to look for a complete solution for a specific problem (\textit{specific solution need}) (10.9\%), intermediate users look for these type of specific complete solutions less (5.3\%) and their questions revolve more around issues that happen while training (10.5\%), looking for documentation (10.5\%), API usage (10.5\%), and understanding the model (13.2\%) or the output of their training/inference (10.5\%). The share of challenges for converting the models is also higher for intermediate users (10.5\%) compared to beginner users (1.5\%) since intermediate users are more involved with modifying models and making custom models.\\
\subsubsection{Analyzing the Trend of Challenges Over Time}\label{RQ1-over_time} In order to better understand the changes in the nature of issues in the forum over time, we assigned the creation date of each topic to all the codes extracted from that topic and compared the distribution of the challenges for each month. We observed no discernible trend for any of the categories of issues over time.

\vspace{-0.5em}
\begin{tcolorbox}[colback=green!5,colframe=green!40!black]
\textbf{Findings}
\begin{enumerate}
    \item[\textbf{1)}] The challenges faced by PTM reusers can be grouped into 17 categories, among which challenges regarding \emph{``Model Usage \& Understanding'', and ``Training Pipeline''} are the most prominent categories.
    \item[\textbf{2)}] Needing a solution for a specific scenario, having difficulty interpreting the output of the model in training or inference, and understanding models and their functionalities are the largest subcategories of challenges faced by HF users.
    \item[\textbf{3)}] Lack of familiarity of the user with HF, its libraries and APIs, and lack of programming and theoretical knowledge by the user are responsible for 31.9\% and 26.4\% of the root causes respectively.
\end{enumerate}
\end{tcolorbox}
\vspace{-0.5em}
\subsection{RQ2. \rqtwo}\label{subsec:RQ2}
In this section, we present the result of our analysis of the benefits of the community for 
PTM reuse. It is worth mentioning that apart from the model hub and the forums, HF provides many resources, tools and modes of communication for the community, including a dataset hub \cite{huggingfaceDatasets}, a Space (model demo) hub \cite{huggingfaceSpaces}, educational resources \cite{huggingfaceDocs,huggingfaceTransformersNotebooks,HuggingFaceEducation,GitHubHuggingfaceeducationtoolkit}, optimization libraries \cite{GitHubHuggingfaceaccelerate}\cite{GitHubHuggingfaceoptimum}, integration with cloud service providers \cite{huggingfacePartnershipAmazon}\cite{huggingfaceOptimumNeuron}, model evaluation \cite{GitHubHuggingfaceevaluate,huggingfaceEvaluatemetricEvaluate,huggingfaceModelEvaluator} and inference and deployment tools \cite{InferenceEndpointsBlog}\cite{huggingfaceHostedInference}. 

\subsubsection{Results}\label{subsubsec:RQ2-results} Table \ref{tab:challenges_and_benefits} summarizes the PTMs reuse benefits identified during the qualitative analysis. It is worth noting that these benefits are gleaned from the discussions on the HF forum and include not only the benefits of model reuse but also the perceived benefits of the HF community for PTM reuse.
The codebook containing the descriptions and examples pertaining to each subcategory of benefits is presented in our replication package \cite{replication}. In the rest of this section, we explain the categories and their corresponding subcategories.

The largest category of benefits, which accounts for around half of the benefits \textbf{B1} (50.97\%) are the instances in which the experts in the community (the HF members and the expert users) have provided clarifications and solutions on different topics to the users.
 
A significant portion (27.43\%) of the observed benefits \textbf{B2} stemmed from various collaborations among users and experts, with expert-user collaboration (6.61\%) being particularly prominent. This collaboration typically involved experts seeking additional information from users to address issues effectively. Additionally, user-user collaboration (5.06\%) and expert-expert collaboration (1.95\%) were noted in joint efforts to solve problems. There were also collaborations where other users suggested solutions, e.g., personal experiences (5.45\%) (e.g., personal experiences and temporary workarounds), and when users found solutions to their own questions and shared them on the forum for the community to benefit (5.25\%).

The final prominent category of benefits is the various forms of acknowledgements of help (with or without feedback on the usefulness of the solution), \textbf{B3} accounting for 14.79\% of the entire set of benefits. The remaining categories of benefits on the forum are the topics created to \textit{Call for collaboration}, \textbf{B4} (2.92\%), e.g., to pre-train a model from scratch for a new language; \textit{Announcements} \textbf{B5} (2.33\%), e.g., introducing a new feature or a new community event; and \textbf{B5}, the identification of individuals from various domains beyond Computer Science (CS), e.g., medicine, business, and art (1.56\%).

\vspace{-0.5em}
\begin{tcolorbox}[colback=green!5,colframe=green!40!black]
\textbf{Findings}
\begin{enumerate}
    \item [\textbf{4)}] The expert-provided solutions and clarifications are the biggest benefit of the HF community.
    \item [\textbf{5)}]The community has made many collaborative projects possible.
    \item[\textbf{6)}] Users from fields outside of CS are coming to the community.
\end{enumerate}
\end{tcolorbox}
\vspace{-0.5em}

\section{Quantitative Analysis}\label{sec:quantitative}
In this section, we explain the motivation, and the methodology adopted for the quantitative analysis to answer RQ3.

\subsection{Motivation}\label{subsec:quant-motivation}
The HF model hub is experiencing rapid growth in the number of models available. By the time of our preliminary analysis in February 2023, there were around 121K models on the hub, and this number nearly doubled by the time of data collection in June 2023 and reached 239k in just 4 months and surpassed the triple number (371K models) in October 2023. Our qualitative analysis findings show that a substantial portion of the detected issues stem from users' struggles to comprehend and effectively use these models. This observation prompted our investigation into the characteristics and types of models causing user issues, examining the distribution of model types discussed by users on the forum in comparison to the types available on the hub.
Since the transformers are the dominating subset of available models over HF, we focus on the \textit{Transformers} library.

\subsection{Methodology}\label{subsec:Quant-methodology}
\subsubsection{Data Collection}\label{subsec:Quant-data_collection} The data available for each model was mined using the \emph{huggingface\_hub} \cite{huggingface_hub}, which is a Python wrapper library of the \emph{Hub API endpoints} \cite{huggingfaceEndpoints} provided by HF that retrieves the data in the form of a JSON file (Figure \ref{fig:methodology}, Step B). For each model data retrieved, there is a field \emph{``model\_type''} which shows the type of the base model or the architecture of the model, e.g., BERT, Llama, etc. While \textit{Transformers} supports more than 200 base models according to its documentation \cite{huggingfaceTransformers}, we referred to the \textit{Transformers} source code to determine how many models they actually recognize as base models. The \textit{Transformers} library uses an auto-mapping based on a set of limited model types to determine the configuration needed for running a model and there were 175 model types present at the time of analysis \cite{githubTransformerssrc} (Figure \ref{fig:methodology}, Step C).
\subsubsection{Types of Models Discussed on the Forums} \label{subsubsec:models-mentions-type}
Since analyzing the existence of mentions of all 239K models from the hub on the forum was not feasible, we decided to look for the base model of the models available on the hub. For extracting the types of models mentioned in the topics, we used case-insensitive pattern matching by regular expressions on the whole text of each topic. Since some model names could be used in other contexts and refer to entities other than the model name (e.g., OPT, Clip, Sam, etc.) the first two authors manually verified the matches concerned by that issue. After manual verification, a total of 7,620 model mentions were identified on the forums in the period of study (147 unique model types) (Figure \ref{fig:methodology}, Step 6).
\subsubsection{Types of Models on the Model Hub}\label{subsubsec:models-hub-type}
To make a comparison between the distribution of the model types discussed on the forums and the model types on the hub, we also extracted the base model type of 239K models on the hub. Therefore, we only counted the model types that were valid, i.e., have a base model name matching one of the 175 model types described in Section \ref{subsubsec:models-mentions-type}. Out of the 239,422 models mined from the hub, only 103,092 models were valid (171 unique model types). The remaining models had model types including the null value, specific model types not general enough to be a base model, a repetition of the name of the model being uploaded, the model owner's name, etc. (Figure \ref{fig:methodology}, Step D).

\subsubsection{Model Creation Dates}\label{subsubsec:models-last-modified-date} For analyzing the trend of different model types added on the hub, we needed to get the date on which each model was created on the hub. For this purpose, we mined the date of the first commit of each model on the hub using the \emph{huggingface\_hub} library described above.
\subsubsection{Model Cards Data}\label{subsubsec:model-cards-data} Another available field in the data retrieved, is the information about the model cards. The model card, first introduced by Mitchel et al. in \cite{mitchell2019model} is a concise accompanying document for a trained ML model that outlines models' performance on benchmarks across diverse conditions and intended usage context, aiming to enhance transparent model reporting. To this date, uploading a model card is not mandatory on HF, but the HF team has recently tried to facilitate and encourage uploading model cards by providing several resources for model owners to use \cite{huggingfaceModelCards}, e.g., a model card creation tool \cite{huggingfaceModelcardCreator}, a model card guide book \cite{huggingfaceModelCardGuidebook}, and model card templates \cite{github_hubtemplatesmodelcard}\cite{huggingfaceAnotatedModelCard} in 2022. We calculated the ratio of models on the hub that had a non-empty model card field, to analyze if the introduction of the tool and guides had an effect on the number of models with card on the hub.

\subsection{RQ3. \rqthree}\label{subsec:RQ3}

\begin{figure*}[t]
    \centering
    \includegraphics[width=0.95\textwidth]{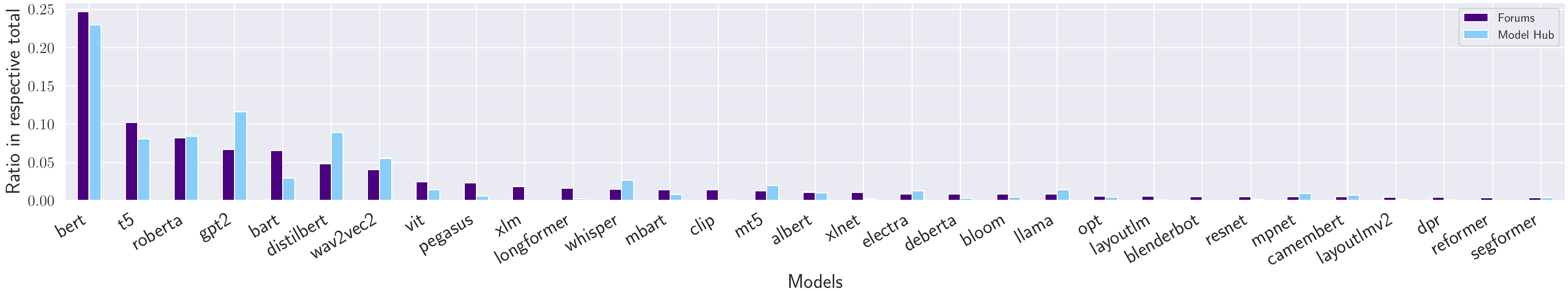}
    \caption{\centering The ratio of the top-31 models in the forums and the model hub}
    \label{fig:modeltypes}
    \vspace{-2em}
\end{figure*}

\subsubsection{Model Types on the Forums and on the Hub}\label{RQ3-types}
The comparison for the distribution of model types in the hub and mentions of the model types in the forums for the 31 most mentioned models on the forum that make up 90\% of the total mentions is depicted in Figure \ref{fig:modeltypes}. \emph{BERT} stands out as the most popular model, both on the hub and among users on the forums. The top-4 model types on the forums and on the hub (responsible for 50\% of each dataset) are the same except for one model type. \emph{GPT2}, and \emph{RoBERTa}, a variation of \emph{BERT} are among the top-4 model types in both groups. \emph{DistilBERT} and \emph{T5} are the other model types in the top-4 model types on the forum and on the hub respectively. One interesting finding is that the most popular models on both the forums and the hub are BERT-based models, i.e., DistilBERT, RoBERTa, and BERT itself. Considering that the Shapiro-Wilk \cite{shapiro1965analysis} test rejected the normality of the distribution of the data, we used Spearman's rank, a non-parametric measure, to calculate the correlation coefficients between the distribution of models on the hub and on the forums, which resulted in a coefficient of 0.702 (\textit{p-value}$\leq$0.01), showing a strong positive correlation.

We also extracted the intersection of the usernames of model providers on the hub and the users active on the forums. There are a total of 73,664 usernames of model providers and 9,299 usernames on the forums. Only 2,299 of the model providers are also present on the forums (3.1\%). Only 158 of these usernames have provided \textit{accepted answers}, among which 54 are internal providers (members of the HF organization).

\subsubsection{Model Types over Time}\label{RQ3-over_time}
We examined the temporal distribution of model types on a monthly basis, utilizing the creation month of each model as well as the creation month of corresponding forum topics as the dates associated with each model/topic (as outlined in Section \ref{subsubsec:models-last-modified-date})

We analyzed growth trends of various models on both forums and the hub. One more time, using the Shapiro-Wilk test, we could reject the normality of the ditribution for any model type, therefore we calculated Spearman's rank correlation coefficients. The analysis covered a 35-month window from July 2020 to May 2023. The results are detailed in Table \ref{tab:spearman_RQ3}.

\begin{table}[t]
\centering
\caption{The Results of the Spearman Correlation Coefficient for the Distribution of on the Hub and Forums}
\vspace{-1em}
\label{tab:spearman_RQ3}
\begin{tabular}{lrr}
\toprule
\textbf{Model} & \multicolumn{1}{l}{\textbf{Correlation Coefficient}} & \multicolumn{1}{l}{\textbf{p-value}} \\ \hline
BART           & -0.6874                                              & \textit{p\textless{}0.05} \\
Pegasus        & -0.5315                                              & \textit{p\textless{}0.05} \\
BERT           & -0.4663                                              & \textit{p\textless{}0.05} \\
RoBERTa        & -0.4234                                              & \textit{p\textless{}0.05} \\
ALBERT         & -0.362                                               & \textit{p\textless{}0.05} \\
ViT            & 0.7655                                               & \textit{p\textless{}0.05} \\
Whisper        & 0.8831                                               & \textit{p\textless{}0.05} \\
LLaMa          & 0.9001                                               & \textit{p\textless{}0.05} \\ 
\bottomrule
\end{tabular}
\vspace{-2em}
\end{table}

In the analysis, significant correlations (\textit{p-value$\leq$0.05}) between some models uploaded on the hub and their corresponding mentions on the forums were observed, among which 8 models belonged to both the top 90\% of the hub and the top 90\% of forum models. Notably, the \textit{BERT} family (BERT, RoBERTa, and ALBERT), BART, and Pegasus exhibited a negative correlation, while the rest displayed a positive correlation. The trends in forum mentions and hub uploads differed for BERT, RoBERTa, Pegasus, and BART (decreasing on the forums but increasing on the hub). BART had the strongest negative correlation (-0.68), whereas Llama showed the strongest positive correlation (0.9). The intense increase in Llama's popularity is expected, considering that Llama is a new Large Language Model (LLM) by Meta which is lighter in size compared to rivals like GPT and was recently released with a commercial license\cite{touvron2023llama}. Whisper and ViT also exhibited strong positive correlations, with models uploaded on the hub and mentions on the forums, on the rise for these two models.

\subsubsection{Models Cards}\label{RQ3-modelcards} 
\begin{figure}
    \centering
    \includegraphics[width=\columnwidth]{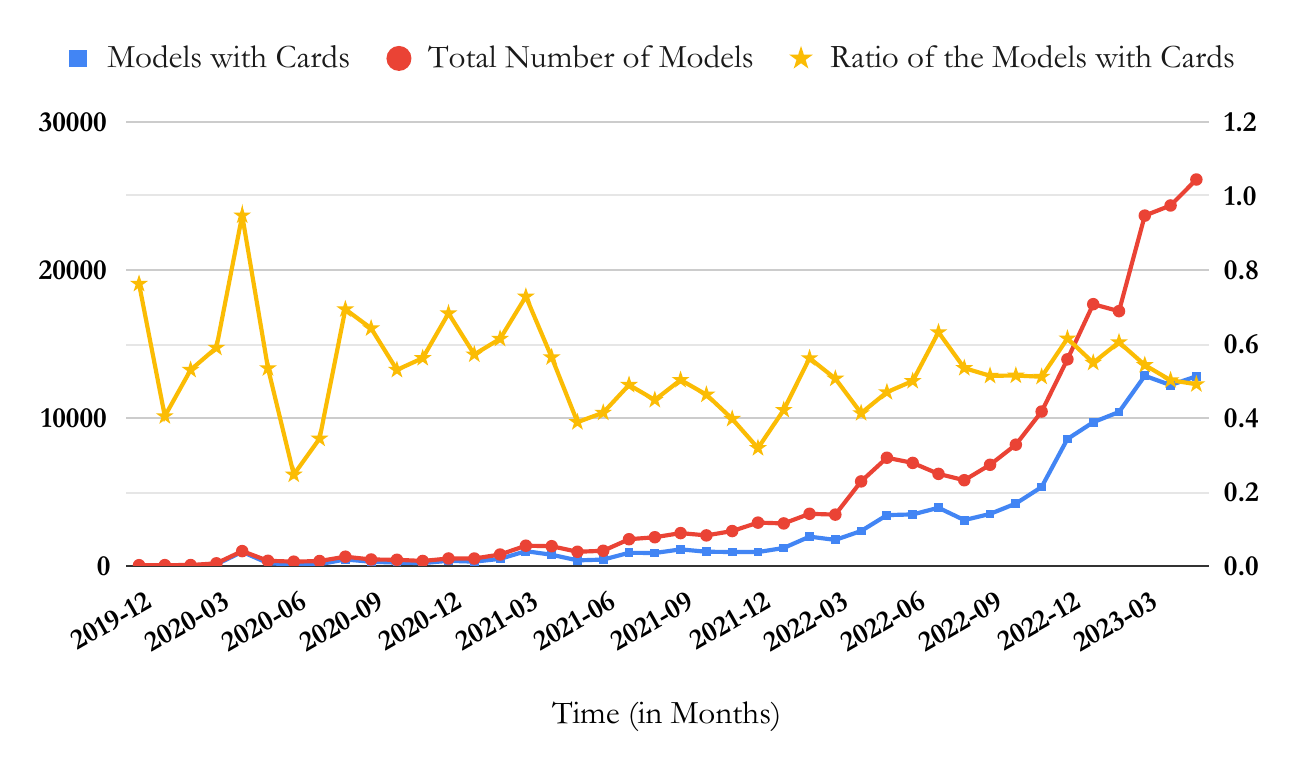}
    \caption{\centering The total number of models and the number of models with model cards uploaded on the hub over time (left axis). The ratio of the models with model cards over time (based on the model creation date) (right axis)}
    \label{fig:models_cards_time}
    \vspace{-1.75em}
\end{figure}

From the 239,422 models mined, only 125,419 (52.38\%) had tags for model cards. The trend of the total models on the hub and models having cards over time (in months, based on the creation date) is reported in Figure \ref{fig:models_cards_time} (left axis). This chart also depicts the ratio of the models in each month that have a model card (right axis).

While the total number of models added on the hub has been increasing at an extremely high pace (26k models created in May 2023), the percentage of the models with model cards has not increased significantly and converges to stagnate below 0.6, almost never passing this ratio.

\begin{tcolorbox}[colback=green!5,colframe=green!40!black]
\textbf{Findings}
\begin{enumerate}
    \item[\textbf{7})] BERT is the most discussed and uploaded model on the forums and the hub and has kept this position over time. T5, RoBerta, and GPT2 are also popular.
    \item[\textbf{8})] There is a negative correlation between the trend of some models provided on the hub and the trend of discussion about these models. Some models like Llama show increasing trends in both.
    \item[\textbf{9})] Despite the introduction of tools to facilitate the model cards and model documentation, the share of model cards on the hub has not improved over time.
\end{enumerate}
\end{tcolorbox}
\vspace{-1em}

\section{Discussion}\label{sec:discussion}

In this section, we discuss our findings for our RQs and compare them with the literature. We also provide suggestions for different parties in the PTM reuse community.

\textbf{Support for beginner users.}
The most prominent subcategories of challenges faced by the users (Finding 2) suggest a need for improved tutorial examples and model guidance, especially for beginner users.

As mentioned in Section \ref{RQ1-expertise},``Beginners'' is the largest category of discussions in the community and there are about seven times more discussions in the ``Beginner'' compared to the ``Intermediate'' category. Moreover, based on the root cause analysis, we found that in 26.4\% of the issues, the problem can be attributed to users lacking sufficient theoretical and programming knowledge.

For the beginners' issues, the most prevalent type of identified questions were instances of users asking a solution for a specific task. In these issues, beginner users were looking for pointers and guidance as a starting point and did not know what model to choose. An example of these issues is seen in \textit{T11271: ``How can I perform a Question-Answering system that returns Yes or No to my questions?[..]Is it possible to do this?[..]Is there any good model to do this?''} \cite{YesNo}. This manifests and highlights the challenge of model search reported by Jiang et al. \cite{jiang2023empirical}; in which the participants pointed out not being able to find the model they need. We also found a relatively large proportion of challenges that users face in finding and following the tutorials and official examples that are provided by \hf (4.6\% of the total). There still seems to be an inadequacy of guidance for people on where to start.\\
\vspace{-1.5em}
\begin{tcolorbox}
[colback=blue!5,colframe=blue!40!black]
\textbf{\faLightbulbO} We recommend that PTM platforms provide dedicated documentation and more examples for specific tasks. They could also provide a platform for users and model providers to share their examples for enhancing the community contribution. Developing a tool similar to HuggingGPT \cite{shen2023hugginggpt} that suggests models, datasets, and hardware requirements for specific tasks can also help the beginner users.
\end{tcolorbox}
On the one hand, we observed onboarding \textbf{C14} issues in our study and on the other hand, in the benefits, we saw newcomer users from fields other than CS (Finding 6). Based on two large-scale surveys by \hf in two consecutive years in 2020 and 2021 \cite{HFCommunitySurvey2020}\cite{HFCommunitySurvey2021}, the share of users having professions other than those related to CS (NLP researcher, data scientist, ML engineer, software engineer and professor) increased by 40\% in a year (5.65\% to 7.96\% of the respondents). 

\begin{tcolorbox}
[colback=blue!5,colframe=blue!40!black]\textbf{\faLightbulbO} We suggest that PTM platform providers simplify the process of PTM reuse for users without adequate knowledge of PTMs, e.g. by suggesting models for specific tasks and providing simplified APIs and step-by-step guides for setting up the environments and training and inference of models.
\end{tcolorbox}
          
\textbf{Model documentation.} The issues regarding model understanding, one of the most prominent groups for both intermediate and beginner users, further underlines the lack of theoretical knowledge and model documentation on both sides of PTM usage, user and provider. This finding is also in line with three challenges in PTM reuse reported by Jiang et al. \cite{jiang2023empirical}, i.e., (1) discrepancies (between claimed and actual performances of models), (2) model application (difficulty in understanding the correct usage of a model for new users), and (3) missing details (in model registries). This challenge, which can also be seen to some part in our documentation issues, could affect the reproducibility of models \cite{jiang2023empirical}.
The second large subcategory of challenges happens when users have a hard time comprehending the output of the model which could point to users' lack of enough theoretical and implementation expertise about DL to be able to troubleshoot, debug, and justify the unexpected results. This issue can also in part be due to the discrepancies in model details. Our Finding 9 also shows another aspect of this issue that despite the tool provided by HF, the quantity of the model cards has not improved over time. The lack of model documentation in quantity leads to more problems and confusion for the user when it comes to model understanding and selection. The multitude of models on the hub with its exponential growth can exacerbate this selection process. Since all model hubs, including HF, are under-exploited \cite{you2022ranking}, the majority of models on the hubs are either superfluous or unnoticed. 
\begin{tcolorbox}
[colback=blue!5,colframe=blue!40!black]\textbf{\faLightbulbO} Enforcing the model card for a set of fields could eradicate the issue of missing information about models to some extent. We also suggest that PTM platforms start taking steps towards verifying the information provided in the model cards to help with users' model understanding and selection and reduce the discrepancies observed in model cards. Providing a simplified interface for searching within the model documentation (model card) can also benefit the users.
\end{tcolorbox}

\textbf{Documentation for platform libraries and tools.}
Our results indicated that many users still face issues in model training, inference, deployment, and API usage issues (in total 19.94\% of challenges) (Finding 1). The training and inference steps are mainly performed using the APIs and tools provided by the platform. Moreover, our root cause analysis showed that the lack of familiarity of the user with HF libraries and APIs and lack of documentation by HF accounted for 31.9\% and 12.5\% of issues respectively (Finding 3 and Table \ref{tab:rootcause}). These findings indicate that despite all the tutorials and documentation provided, users are not fully familiar with how to utilize the tools provided by the platform to use PTMs. This could be because the multitude of libraries provided by HF makes mastering all of them hard for the user, or there is a mismatch between the level of difficulty of the APIs and the average level of the users. In any case, this calls for further investigation in regard to the actual reason. 

\textbf{Involvement of model providers.}
While the hub has made many collaborations between experts and users and within these groups possible (Finding 5), the majority of model providers are not present on the forums and their absence could indicate a lack of involvement. We observe that some models (mostly of the BERT family) are being discussed less and yet the number of models being uploaded on the hub is still increasing for them. While the decrease in the discussions about these models could be due to the fact that these models were mostly introduced before 2020 and therefore are well-established, the trend is still worth noting for model providers.
\begin{tcolorbox}
[colback=blue!5,colframe=blue!40!black]\textbf{\faLightbulbO} 
We suggest HF to encourage PTM providers to consider the trends in the models' usage and try to provide models that are consistent with the community's needs.
\end{tcolorbox}

\section{Related Work}\label{sec:relatedwork}
\subsection{PTM Reuse and Model Hubs}
Pan et al. \cite{pan2022empirical} studied the GitHub repositories of popular models supported by HF transformers, analyzing reported bugs to create a detailed taxonomy, and uncovering root causes and implications for model reuse. Our study scope encompasses issues beyond bugs, as we analyze the whole discussion forums. We also look at model reuse for a larger variety of models rather than selected NLP models. McMillan-Major et al.\cite{mcmillan2021reusable} propose documentation for language datasets and NLG models based on HuggingFace Hub and GEM benchmark. Jiang et al. \cite{jiang2022empirical} conducted an empirical study assessing artifacts and security features across 8 model hubs. Their work highlighted potential threat models and demonstrated the inadequacy of current defenses in securing PTMs. In a recent study, Jiang et al. \cite{jiang2023empirical} conducted interviews with 12 HF practitioners, delving into PTM reuse practices and challenges within the HF ecosystem, and proposing a decision-making process. They also analyzed 63k PTM packages on HF, constructing a dataflow model for PTM creation and distribution. Our study differs from theirs in both study type and scale. Their participants were all of intermediate and expert level, but most of our topics belonged to the beginners' category. Our results confirm and complement their detected challenges. Shen et al.\cite{shen2023hugginggpt} introduced a tool, \textit{HuggingGPT}, to manage multiple HF models and find solutions according to a specific task of the user's needs. You et al. \cite{you2022ranking} proposes a novel approach for ranking PTMs on the PTM hub in two steps: ranking PTMs based on a transferability metric, and fine-tuning the top K-ranked PTMs using a Bayesian procedure to meet the requirements of downstream applications. Costano et al. \cite{castano2023exploring} analyzed the carbon-footprint reports available for the models on the HF hub and reported that the proportion of carbon emissions-reporting models on HF has stalled and analyzed the correlation between carbon emission and various model attributes. Ait et al. \cite{ait2023hfcommunity} also introduced a relational database of all the available data for each repository on HF (i.e., model, dataset, or space) called HFCommunity, and provided some metrics of the HF repositories based on that. In another recent work, Jiang et al. \cite{jiang2023exploring} explored the naming conventions and defects in PTM packages in different model hubs including HF, and proposed a method to detect the defects.

 \subsection{Software Communities and Q\&A Platforms}
There are several works that use topic modeling to automatically detect the topics of Q\&A forums. Barua et al. \cite{barua2014developers} utilized Latent Dirichlet Allocation (LDA) on Stack Overflow discussions to analyze the trends, popularity, and relationships of these topics over time. Scoccia et al. \cite{scoccia2021challenges} also used LDA on StackOverflow and GitHub to identify topics related to developing desktop web applications. They analyzed the difficulty of each topic's questions and compared the two platforms. Zhang et al. \cite{tianyi2019RCA} also used qualitative analysis on StackOverflow posts to detect the challenges in developing DL applications. They also identified the root causes.

\section{Threats to the Validity}\label{sec:threats}

\textit{\textbf{Construct Validity:}} 
Relates to the risk associated with measurement errors. The choice of HF Forums may introduce bias in PTM discussions by not fully representing all discussions. To address this, we use the snowball effect to track outside links within the forum discussion. Additionally, we relied on HF's decision to decongest GitHub and keep it dedicated while centralizing discussions in the Forum. Also, we used open-source tools and regular expressions for data mining and cleaning, potentially introducing selection bias. To mitigate this, we employed established tools and methods commonly used in the literature.

\textit{\textbf{Internal Validity:}} These threats pertain to possible alternative interpretations of our results. Our emerged categories might have been susceptible to confirmatory or subjective bias, typical in qualitative studies. To preempt this, we were guided by the literature and conducted multiple rounds of negotiated agreements to achieve a higher kappa score. An underrepresented sample size is a potential threat. To address this, we employed random sampling with a 95\% confidence level and a 5\% margin of error. Then, we continued coding until saturation. As we observed a low rate of accepted answers for the HF forums (0.09 and 0.11 for the whole dataset and our selected samples, respectively), our sample set looks to be representative. The validity of the root cause analysis in this study may be threatened by the small sample size of 72. However, we took measures to mitigate this threat by manually inspecting and analyzing the topics to identify the number of unmarked accepted answers and available instances of resolved challenges in our data. The ratio of the root cause analysis sample to the whole sample is consistent with reported values in similar studies in the literature \cite{tianyi2019RCA}.

\textit{\textbf{External Validity:}}
Although HuggingFace is a well-known platform for reusing PTMs, our research may not apply to other PTM reuse communities. However, we chose HuggingFace because it is a leading PTM hub that offers free models. Moreover, Forums are the primary platform for questions and discussions within the HuggingFace community. While our findings may not generalize to other PTM communities and model hubs, yet, our suggestions and recommendations are relevant to any platform and provide practical steps.

\textit{\textbf{Reliability Validity:}}
To enhance reliability and enable replication, we have included all necessary details for reproducing our study, and our complete replication package is available online~\cite{replication}.

\section{Conclusion}\label{sec:conclusion}
In this study, we performed a large-scale mixed-methods analysis on the HuggingFace platform, which is the biggest open model hub available. As a result of our qualitative coding, we could detect 17 categories of challenges faced by the users in the PTM reuse community, with a lack of guidance and pointers for beginner users, incomprehensible model outputs, and lack of model understanding being the most prominent issues faced by the users, especially beginners. The analysis of the root causes of these challenges also pointed to the user's lack of familiarity with HF and its resources and lack of theoretical and programming knowledge by the user. On the other hand, the community led to multiple benefits for PTM reuse such as many collaborations in the community of users and experts. Our quantitative analysis showed that there is a negative correlation between the trend on the hub and on the forums, for some prevalent models (like BERT) that might indicate a lack of synchronization between model providers and users.
We provided guidelines and recommendations for model providers and PTM providing platforms. 

In future work, we plan to do a user study with different partners in PTM reuse community to investigate the challenges they face and confirm our findings. We also aim to further investigate the quality and quantity of model documentation and its effect on PTM reuse.

\bibliographystyle{IEEEtran}
\bibliography{_bibliography.bib}

\end{document}